\documentstyle[12pt]{article}
\def\bbbc{{\mathchoice {\setbox0=\hbox{$\displaystyle\rm C$}\hbox{\hbox
to0pt{\kern0.4\wd0\vrule height0.9\ht0\hss}\box0}}
{\setbox0=\hbox{$\textstyle\rm C$}\hbox{\hbox
to0pt{\kern0.4\wd0\vrule height0.9\ht0\hss}\box0}}
{\setbox0=\hbox{$\scriptstyle\rm C$}\hbox{\hbox
to0pt{\kern0.4\wd0\vrule height0.9\ht0\hss}\box0}}
{\setbox0=\hbox{$\scriptscriptstyle\rm C$}\hbox{\hbox
to0pt{\kern0.4\wd0\vrule height0.9\ht0\hss}\box0}}}}
\def\bbbr{{\rm I\!R}} %reelle Zahlen
\def\bbbone{{\mathchoice {\rm 1\mskip-4mu l} {\rm 1\mskip-4mu l}
{\rm 1\mskip-4.5mu l} {\rm 1\mskip-5mu l}}}
\def\bbbc{{\mathchoice {\setbox0=\hbox{$\displaystyle\rm C$}\hbox{\hbox
to0pt{\kern0.4\wd0\vrule height0.9\ht0\hss}\box0}}
{\setbox0=\hbox{$\textstyle\rm C$}\hbox{\hbox
to0pt{\kern0.4\wd0\vrule height0.9\ht0\hss}\box0}}
{\setbox0=\hbox{$\scriptstyle\rm C$}\hbox{\hbox
to0pt{\kern0.4\wd0\vrule height0.9\ht0\hss}\box0}}
{\setbox0=\hbox{$\scriptscriptstyle\rm C$}\hbox{\hbox
to0pt{\kern0.4\wd0\vrule height0.9\ht0\hss}\box0}}}}

\def\oplusinf{\mathop{\oplus}}

\begin{document}

\begin{titlepage}
\begin{center}

{\Large COMPLEX STRUCTURES AND}

\vspace{0,3cm}

{\Large THE ELIE CARTAN APPROACH}

\vspace{0,3cm}

{\Large TO THE THEORY OF SPINORS}

\end{center}

\vspace{1cm}

\begin{center}
{\bf\large Michel DUBOIS--VIOLETTE}
\vspace{0.5cm}

Laboratoire de
Physique Th\'eorique et Hautes Energies\\
B\^atiment 211, Universit\'e Paris~XI\\
 91405 ORSAY Cedex, France \\ E-mail: FLAD@QCD.CIRCE.FR
\end{center}
\vspace{1cm}

\begin{center}
Lecture given at the
Second Max Born Symposium\\ ``Spinors, Twistors and
Clifford Algebras"\\ held in Wroc\l aw, Poland, Sept.
24--27, 1992.
\end{center}
\vspace{1cm}
\begin{abstract}
Each isometric complex structure on a
2$\ell$-dimensional euclidean space $E$ corresponds to an
identification of the Clifford algebra of $E$ with the
canonical anticommutation relation algebra for $\ell$ (
fermionic) degrees of freedom. The simple spinors in the
terminology of E.~Cartan or the pure spinors in the one
of C. Chevalley are the associated vacua. The
corresponding states are the Fock states (i.e. pure free
states), therefore, none of the above terminologies is
very good.
\end{abstract}

\vspace{3cm}

\indent {\large L.P.T.H.E. -- ORSAY 92/53}

\end{titlepage}

\section{Introduction}
In this lecture, we will discuss complex structures and
spinors on euclidean space. This is an extension of the
algebraic part of a work [1] describing a sort of
generalization of Penrose and Atiyah--Ward
transformations in $2\ell$ dimension. We shall not
describe this work here, refering to [1], but
concentrate the lecture upon the notion of simple spinor of
E.~Cartan [2] (or pure spinor in the terminology of
C.~Chevalley [3]). Many points of this lecture are
well known facts and, in some sense, this may be
considered as an introductory review. The notations used
here are standard, let us just point out that by an
euclidean space we mean a {\sl real} vector space with a
positive scalar product and by a Hilbert space we mean a
{\sl complex} Hilbert space.

\section{Isometric Complex Structures}

\subsection{Notations}
 Let $E$ be an oriented 2$\ell$-dimensional
euclidean space ($E\simeq \bbbr^{2\ell}$) with a scalar
product denoted by ($\bullet, \bullet$). The dual space $
E^\ast$ of $E$ is also, in a canonical way, an euclidean
space and we again denote its scalar product by ($\bullet,
\bullet$). On the complexified space $E^\ast_c = E^\ast
\otimes \bbbc$ of $E^\ast$ one may extend the
scalar product of $E^\ast$ in two different ways: Either
one extends it by bilinearity and the corresponding
bilinear form will again be denoted by ($\bullet, \bullet$)
or one extends it by sesquilinearity and the corresponding
sesquilinear form will be denoted by $\langle \bullet
\vert \bullet\rangle$. As for any complexified vector
space, there is a canonical complex conjugation $\omega
\mapsto \bar\omega$ on $E^\ast$, (an antilinear
involution), and the connection between the two scalar
products is given by:
$$\langle\omega_1 \vert \omega_2\rangle = (\bar \omega_1,
\omega_2), \qquad \forall \omega_1, \omega_2 \in
E^\ast_c.$$

\subsection{Isometric Complex Structures or Hilbertian
Structures}
Let ${\cal H}(E)$ be the set of isometric complex
structures on $E$ or, which is the same, the set of
orthogonal antisymmetric endormorphisms of $E$, i.e.
$${\cal H}(E) = \{J\in {\rm End} (E) \vert J\in O(E)\
{\rm and}\  J^2=-\bbbone \} =$$
$$= \{J\in {\rm End}(E) \vert J\in O(E)\
{\rm and}\  (X,JY) = -(JX,Y), \quad \forall X,Y\in E\}$$\
Let $J\in {\cal H}(E)$ and define
$$(x+iy) V = xV + yJV, \qquad \forall (x+iy)\in \bbbc,
\qquad \forall V \in E$$
and
$$\langle X\vert Y\rangle_J = (X,Y) - i(X,JY), \qquad
\forall X,Y \in E.$$
Equipped with the above structure, $E$ is a
$\ell$-dimensional Hilbert space which we denote by
$E_J$. For a basis $(e_1,\dots , e_\ell)$ of the complex
vector space $E_J$, ($e_1, \dots, e_\ell , Je_1, \dots ,
Je_\ell$) is a basis of $E$ the orientation of which is
independent of $(e_1, \dots , e_\ell$) but only depends
on $J$. Accordingly, ${\cal H}(E)$ splits in two pieces :
${\cal H}(E) = {\cal H}_+(E) \cup {\cal H}_-(E)$. The
orthogonal group $O(E)$ acts transitively on ${\cal H}(E)$
and the subgroup $SO(E)$ of orientation preserving
orthogonal transformations acts transitively on
${\cal H}_+(E)$ and on ${\cal H}_-(E)$.\\
Thus one has ${\cal H}(E)\simeq O(E) / U(E_J)$ and
${\cal H}_+(E)\simeq SO(E) / U(E_J) \simeq {\cal H}_-(E)$
where $U(E_J)$ is the unitary group of $E_J$ for a fixed
$J\in {\cal H}(E)$ (i.e. $U(E_J)\simeq U(\bbbc^\ell)).$  We
equip ${\cal H}(E), {\cal H}_+(E)$ and ${\cal H}_-(E)$ with
the corresponding manifold structure. In particular,
${\rm dim}_\bbbr{\cal H}(E) = {\rm dim}_\bbbr
{\cal H}_\pm (E) = \ell(2\ell -1) - \ell^2 = \ell(\ell
-1).$

\subsection{Identification of Dual Spaces}
The dual Hilbert space of $E_J$ can be identified with
the Hilbert subspace $\Lambda^{1,0}E_J^\ast$ of
$E^\ast_c$ defined by
$$\Lambda^{1,0}E^\ast_J = \{\omega\in E^\ast_c \vert
\omega\circ J = i\omega\}.$$
One verifies easily that $\Lambda^{1,0}E_J^\ast$ is
maximal isotropic in $E^\ast_c$ for ($\bullet ,
\bullet$) or, which is the same, that
$\Lambda^{1,0}E_J^\ast$ is orthogonal to its conjugate
$\overline{\Lambda^{1,0}E^\ast_J}=\Lambda^{0,1}E^\ast_J$
in $E^\ast_c$ for $\langle \bullet\vert\bullet\rangle$
(i.e. $E^\ast_c$ is the hilbertian direct sum
$\Lambda^{1,0}E^\ast_J \oplus \Lambda^{0,1} E^\ast_J$).\\
Conversely if $F \subset E^\ast_c$ is a maximal isotropic
subspace for ($\bullet , \bullet$), then there is a
unique $J\in {\cal H}(E)$ such that
$F=\Lambda^{1,0}E^\ast_J$. It follows that ${\cal H}(E)$
identifies with a complex algebraic submanifold of the
grassmannian $G_\ell(E^\ast_c)$ of $\ell$-dimensional
subspaces of $E^\ast_c$, $(G_\ell(E^\ast_c)\simeq
G_{\ell,2\ell}(\bbbc))$. In particular, ${\cal H}(E)$ is
a compact K\"ahler manifold of complex dimension
$\frac{\ell(\ell-1)}{2}$ and its K\"ahler metric is given
by $ds^ 2=\frac{1}{4} {\rm tr} ((dP^{1,0}_J)^2)$ where
$P^{1,0}_J$ is the hermitian projector of $E^\ast_c$ on
$\Lambda^{1,0}E^\ast_J$. Notice that one has
$\overline{P^{1,0}_J} = P^{0,1}_J = \bbbone - P^{1,0}$.\\
Furthermore $\Lambda^{1,0}E^\ast_J$ is the fibre at $J\in
{\cal H}(E)$ of a holomorphic hermitian vector bundle of
rank $\ell$ over ${\cal H}(E)$ which we denote by
$\Lambda^{1,0}E^\ast$.\\
Finally notice that one has the hilbertian
sum identifications
$$\Lambda^kE^\ast_c = \oplusinf_{r+s=k} \Lambda^{r,s}
E^\ast_J, \qquad \forall J \in {\cal H}(E)$$
where $\Lambda^{r,s}E^\ast_J =
\Lambda^r(\Lambda^{1,0}E^\ast_J) \otimes \Lambda^s
(\overline{\Lambda^{1,0}E^\ast_J})$, (here the tensor
product is over $\bbbc$).We denote by $P^{r,s}_J$ the
corresponding hermitian projectors.

\subsection{Examples}
One has ${\cal H}_+(\bbbr^2) = \{I_+\}, {\cal
H}_+(\bbbr^4) = \bbbc P^1, {\cal H}_+(\bbbr^6) = \bbbc
P^3$ and, as will be shown below, ${\cal
H}_+(\bbbr^{2\ell}) \subset \bbbc P^{2^{\ell-1}-1}$ but the
inclusion is strict for $\ell\geq 4$ as it follows by
comparison  of the dimensions.

\subsection{Hodge duality}
On $\Lambda E^\ast$ there is a linear involution, $\ast$,
defined by $\ast (\omega^1 \wedge \dots \wedge
\omega^p)=\omega^{p+1} \wedge \dots \wedge \omega^{2\ell}$
for any positively oriented orthonormal basis $(\omega^1,
\dots ,\omega^{2\ell})$. One extends this involution by
linearity to $\Lambda E^\ast_c$. One has the following
lemma.\\

\noindent {\bf Lemma.}
 {\sl Let $\Omega$ be an element of
$\Lambda^\ell E_c^\ast$. Then one has $\Omega + i^\ell
\ast \Omega =0$, (resp. $\Omega-i^\ell \ast \Omega = 0$),
if and only if $P^{0,\ell}_J \Omega = 0$, $\forall J\in
{\cal H}_+(E)$, (resp. $\forall J\in {\cal H}_-(E)).$}\\

\noindent For $\ell = 2$ (i.e. in dimension 4), this is the
basic algebraic lemma for the Penrose--Atiyah--Ward
transformation.

\section{The Clifford algebra as C.A.R. algebra}
\subsection{Definition}
We define the Clifford algebra ${\rm Cliff}(E^\ast)$ to
be the complex associative $\ast$-algebra with a unit
$\bbbone$ generated by the following relations
$$[\gamma(\omega_1), \gamma(\omega_2)]_+ = 2(\omega_1,
\omega_2) \bbbone\ {\rm and}\  \gamma(\omega)^\ast =
\gamma(\omega)\ {\rm for}\ \omega, \omega_i\in E^\ast.$$
The $\gamma(\omega), \omega \in E^\ast$, are hermitian
generators and $\gamma : E^\ast \rightarrow {\rm Cliff}
(E^\ast)$ is an injective $\bbbr$-linear mapping. One
extends $\gamma$ as a $\bbbc$-linear mapping, $\gamma :
E^\ast_c \rightarrow {\rm Cliff}(E^\ast)$, by setting
$\gamma(\bar \omega)=\gamma(\omega)^\ast$.

\subsection{Complex structures and the C.A.R. algebra}
Let $J\in {\cal H} (E)$ be given. The algebra ${\rm
Cliff}(E^\ast)$ is generated by the $\gamma(\omega)$
with  $\omega\in \Lambda^{1,0}E^\ast_J$ and their
adjoints $\gamma(\omega)^\ast = \gamma (\bar\omega)$. In
terms of these generators the relations read
$$[\gamma(\omega_1), \gamma(\omega_2)]_+=0\ {\rm and}\
[\gamma(\omega_1)^\ast, \gamma(\omega_2)]_+ =
\langle\omega_1 \vert \omega_2\rangle \bbbone, \quad
\forall \omega_i \in \Lambda^{1,0}E^\ast_J.$$
These are the defining relations of the algebra of
canonical anticommutation relations (C.A.R. algebra) for
$\ell$ (fermionic) degrees of freedom. Thus each $J\in
{\cal H}(E)$ corresponds to an identification of the
Clifford algebra with the C.A.R. algebra. Furthermore, the
action of the orthogonal group $O(E)$ on ${\cal H}(E)$
corresponds to the Bogolioubov transformations. One has,
as well known, ${\rm Cliff} (E^\ast) \simeq
M_{2^\ell}(\bbbc)$.

\section{Spinors and Complex Structures}
\subsection{Definition}
We define a space of spinors associated to $E$ to be a
Hilbert space $S$ carrying an irreducible
$\ast$-representation of ${\rm Cliff}(E^\ast)$. The
spinors are the elements of $S$. Since ${\rm
Cliff}(E^\ast)$ is isomorphic to $M_{2^\ell}(\bbbc)$, $S$
is isomorphic to $\bbbc^{2^\ell}$ and the representation is
an isomorphism. We shall identify ${\rm Cliff} (E^\ast)$
with the image of this representation.

\subsection{The simple spinors of E. Cartan}
Let $\psi\in S$ with $\psi\not= 0$ and set $I_\psi =
\{\omega\in E^\ast_c \vert \gamma(\omega) \psi = 0\}$. If
$\omega_1$ and $\omega_2$ are in $I_\psi$, one has
$[\gamma(\omega_1), \gamma(\omega_2)]_+ \psi = 2
(\omega_1, \omega_2)\psi = 0$, so $I_\psi$ is an
isotropic subspace of $E^\ast_c$ for ($\bullet,
\bullet$).\\
If $I_\psi$ is maximal isotropic, i.e. if ${\rm dim}
(I_\psi)=\ell$, then $\psi$ is called a {\sl simple
spinor} by E. Cartan [2] or a {\sl pure spinor} by C.
Chevalley [3]. We denote by ${\cal F}$ the set of these
spinors and by $P({\cal F})$ the corresponding algebraic
submanifold of $P(S)=\bbbc P^{2^{\ell}-1}$, (i.e. $P({\cal
F})$ is the set of directions of simple spinors).\\
For $\psi \in {\cal F}, I_\psi=I_{\lambda\psi}\quad
\forall \lambda\in \bbbc \backslash \{0\}$, so the
maximal isotropic subspace $I_\psi$ of $E^\ast_c$ does
only depend on the direction $[\psi]\in P({\cal F})$ of
$\psi$.
On the other hand we know that there is a unique $J\in
{\cal H}(E)$ such that $I_\psi=\Lambda^{1,0}E^\ast_J$. It
follows that one has a mapping of $P({\cal F})$ in ${\cal
H}(E)$ which is in fact an isomorphism of complex
manifolds. In the following, we shall identify these
manifolds, writing $P({\cal F})={\cal H}(E)$.

\subsection{The natural line bundle}
The restriction to $P({\cal F}) = {\cal H}(E)$ of the
tautological bundle of $P(S)$ is a holomorphic hermitian
vector bundle of rank one, $L$, over ${\cal H}(E)$. One
has $L = {\cal F} \cup \{{\rm the\  zero\  section}\}$.\\
As holomorphic hermitian line bundles over ${\cal H}(E)$,
one has the following isomorphisms, see in [1] :
$\Lambda^{\ell,0}E^\ast \simeq L\otimes L$ and
$\Lambda^{\frac{\ell(\ell -1)}{2}} T^\ast {\cal H} (E) =
L^{\otimes\ 2 (\ell -1)}$.

\subsection{Semi-spinors and simple spinors}
To the action of $SO(E)$ on $E^\ast$ corresponds a
linear representation of its covering ${\rm Spin}(E)$ in
$S$. Under this representation, $S$ splits into two
irreductible components $S=S_+ \oplus S_-$ with ${\rm
dim} S_+ = {\rm dim} S_- = 2^{\ell-1}$. The elements of
$S_+$ and $S_-$ are called semi-spinors. One the other
hand $P({\cal F})={\cal H}(E)$ splits into two transitive
parts under the action of $SO(E)$, ${\cal H}(E) = {\cal
H}_+(E) \cup {\cal H}_-(E)$. It follows that ${\cal F} =
{\cal F}_+ \cup {\cal F}_-$ with ${\cal F}_\pm = {\cal F}
\cap  S_\pm$ and  (with an eventual relabelling in the
$\pm$)  $P({\cal F}_\pm) = {\cal H}_\pm (E)$. In other
words ${\cal F}$ consists of semi-spinors. It turns out
that for $\ell\leq 3$ all non vanishing semi-spinors are in
${\cal F}$ (i.e. ${\cal F}_\pm = S_\pm\backslash \{0\})$
but for $\ell \geq 4$ the inclusions ${\cal F}_\pm \subset
S_\pm\backslash \{0\}$ are strict inclusions. For $\ell\geq 4$ ${\cal
H}_+(E)$ is no more a projective space.

\section{Fock States and Simple Spinors}
\subsection{States on algebras}

Let ${\cal A}$ be an associative complex $\ast$-algebra
with a unit $\bbbone$. We recall that a state on ${\cal
A}$ is a linear form $\phi$ on ${\cal A}$ such that $\phi
(X^\ast X) \geq 0,\quad \forall X\in {\cal A}$ and
$\phi(\bbbone)=1$. The set of all states on ${\cal A}$ is
a convex subset of the dual space ${\cal A}^\ast$ of
${\cal A}$. The extreme points of this convex subset
(i.e. which are not convex combinations of two distinct
states) are called pure states. To the states on ${\cal
A}$ correspond cyclic $\ast$-representations of ${\cal
A}$ in Hilbert space via the G.N.S. construction; pure
states correspond then to irreducible representations.\\
Coming back to the case ${\cal A} = {\rm Cliff}(E^\ast)$,
we see that to each spinor $\psi\not=0$ corresponds  a
state $X \mapsto \frac{\langle \psi \vert
X\psi\rangle}{\parallel \psi\parallel^2}$ (its direction
) which is a pure state leading to an irreducible, or
simple, representation. This is why the terminology of
C.~Chevalley or E.~Cartan to denote the elements of
${\cal F}$ is somehow misleading. What characterizes the
elements of ${\cal F}$ is that the corresponding states
(i.e. elements of $P({\cal F}) = {\cal H}(E))$ are Fock
states or free states on ${\rm Cliff} (E^\ast)$ (see
below); thus the name Fock spinors or free spinors would
be better.

\subsection{Fock states on the Clifford algebra}
First of all it is clear from above that the elements of
${\cal F}$ are all possible vacua corresponding to the
identifications of ${\rm Cliff}(E^\ast)$ with the C.A.R.
algebra. It is well known that given a vacuum, the
vacuum expectation values factorize and only depend on
the ``two-point functions" i.e. on the vacuum expectation
values of $\gamma(\omega_1)\gamma(\omega_2)$ for
$\omega_i\in E^\ast$, (this is the very property of the
free states).\\
More precisely, a {\sl Fock state}, (see for instance
[4]),  on ${\rm Cliff}(E^\ast)$ is a pure state $\phi$
satisfying the following (Q.F.) property:

$$(Q.F.) \left\{ \begin{array}{l}
\phi(\gamma(\omega_1) \dots \gamma(\omega_{2n+1}))=0\\
\phi(\gamma(\omega_1) \dots \gamma(\omega_{2n})) =
\sum_{k=2}^{2n}(-1)^k\phi(\gamma(\omega_1)\gamma(\omega_k)).
\phi(\gamma(\omega_2).\buildrel {\rm k \atop ^\vee}
\over .. \gamma(\omega_{2n}))
\end{array}
\right. $$
 for $\omega_i\in E^\ast$, (where $\buildrel {\rm k \atop
^\vee} \over .$ means omission of the k$^{\rm th}$ term).
{}From (Q.F.) one sees that $\phi$ is determined by the
$\phi(\gamma(\omega_1)\gamma(\omega_2)) =
h(\omega_1,\omega_2)+i\sigma (\omega_1, \omega_2),$
 $\omega_i \in E^\ast$, where $h$ and $\sigma$ are real
bilinear forms. The defining relations of ${\rm
Cliff}(E^\ast)$ implie that
$h(\omega_1,\omega_2)+h(\omega_2, \omega_1)=2(\omega_1,
\omega_2)$ and $\sigma (\omega_1,\omega_2)+\sigma(
\omega_2, \omega_1)=0$. The positivity of $\phi$ is
equivalent to $\phi(\gamma(\omega_1+i\omega_2) \gamma
(\omega_1-i\omega_2))\geq 0$ which is equivalent to
$h(\omega_1, \omega_2)=(\omega_1, \omega_2)$ and
$\sigma(\omega_1,\omega_2)= (A\omega_1, \omega_2)=-
(\omega_1, A\omega_2)$ with $\parallel A \parallel \leq
1$. By polar decomposition, $A=J\vert A\vert$ with $J\in
{\cal H} (E)$ and\\$\vert A\vert \geq 0$ ($\parallel\vert
A\vert \parallel\geq 1$). Then, $\phi$ is pure if and
only if $\vert A\vert = 1$. Therefore, $\phi$ is a Fock
state iff. it satisfies (Q.F.) and
$\phi(\gamma(\omega_1)\gamma(\omega_2)) =
(\omega_1,\omega_2)+ i(J\omega_1,\omega_2),\quad \forall
\omega_i\in E^\ast$,  with $J\in {\cal H}(E)$. Thus, the
Fock states are parame\-trized by ${\cal H}(E)=P({\cal
F})$ and, in fact, the set of Fock states is $P({\cal F})$;
indeed if $\psi\in {\cal F}$ is such that
$I_\psi=\Lambda^{1,0}E^\ast_J$ then one has
$$\frac{\langle\psi\vert
\gamma(\omega_1)\gamma(\omega_2)\psi\rangle}{\parallel
\psi\parallel^2} =
(\omega_1,\omega_2)+i(J\omega_1,\omega_2), \quad \forall
\omega_i\in E^\ast$$
and (Q.F.) is satisfied.
\section{Spinors and Fock Space Constructions}
The standard construction of the Fock space for the
C.A.R. algebra implies that, for each $J$, $S$ is
isomorphic to
$$\oplusinf_n \Lambda^{0,n}E^\ast_J.$$
However, there is the vacuum, namely an element of $L_J$,
which is hidden here.\\
In fact, one has an isomorphism $\Phi$ of hermitian
vector bundles over ${\cal H}(E)$ from
$$\oplusinf_n \Lambda^{0,n}E^\ast\otimes L$$
onto the trivial bundle with fibre equal to $S$, such that
$$\Phi(\omega\wedge\varphi) = \frac{1}{\sqrt{2}}
\gamma(\omega)\Phi(\varphi), \quad \forall
\omega\in\Lambda^{0,1}E^\ast_J\
{\rm and}\ \forall
\varphi \in \oplusinf_n\Lambda^{0,n} E^\ast_J\otimes
L_J.$$
More precisely one has the following
$$ \left.  \begin{array}{lll}

\Phi_J : \oplusinf_p\Lambda^{0,2p} E^\ast_J \otimes L_J
\simeq S_+ & ({\rm resp.}\ S_-)\\
\Phi_J :\oplusinf_p \Lambda^{0,2p+1} E^\ast_J\otimes L_J
\simeq S_- & ({\rm resp.}\ S_+) \end{array}
\right\} \forall J\in {\cal H}_+ (E)\ ({\rm resp.} {\cal
H}_-(E))$$
which gives the identification of semi-spinors.

\section{Bundles of Complex Structures}

Let $M$ be a $2\ell$-dimensional oriented riemannian
manifold. The tangent space $T_x(M)$ at $x\in M$ is an
oriented $2\ell$-dimensional euclidean space so one can
consider the complex manifold ${\cal H}(T_x(M))$ as
above. ${\cal H}(T_x(M))$ is
the fiber at $x\in M$ of a bundle ${\cal H}(T(M))$ on M
which we call the {\sl bundle of isometric complex
structures over} $M$. This bundle is associated to the
orthonormal frame bundle so there is a natural connection
on it coming from the Levi--Civita connection of $M$. On
${\cal H}(T(M))$, there is a {\sl natural almost complex
structure} defined by the following construction. Let
$J_x\in {\cal H} (T_x(M))$, then by horizontal lift,
$J_x$ defines a complex structure on the tangent
horizontal subspace at $J_x$; on the other hand the
tangent vertical subspace at $J_x$ is the tangent space
to the complex manifold ${\cal H}(T_x(M))$ so it is
naturally a complex vector space, so by taking the direct
sum one has a complex structure on the tangent space to
${\cal H}(T(M))$ at $J_x$ and finally, ${\cal H}(T(M))$
becomes an almost complex manifold. It is easy to show
that the almost complex manifold ${\cal H}(T(M))$ only
depends on the conformal structure of $M$. In particular,
the almost complex structure of ${\cal H}(T(M))$ is
integrable, i.e. ${\cal H}(T(M))$ is a complex manifold,
whenever $M$ is conformally flat. The Penrose and the
Atiyah-Ward transformations are obtained, in the
four-dimensional case, by lifting to ${\cal H}(T(M))$
various objects living on $M$ (see in [1]).\\
Let us end this lecture by noticing that the complex
manifold ${\cal H} (T(S^{2\ell}))$ identifies with the
complex manifold ${\cal H}(\bbbr^{2\ell+2})$ of
isometric complex structures on the euclidean space
$\bbbr^{2\ell +2}$, [1]. So, in particular, by restriction
to the positively oriented complex structures one has
${\cal H}_+(T(S^4)) = {\cal H}_+(\bbbr^6) = \bbbc P^3$.

 \end{document}